\documentclass[10pt,conference]{IEEEtran}
\IEEEoverridecommandlockouts
\usepackage{algpseudocode}
\usepackage{amsmath}  
\usepackage{CJK}
\usepackage{array}
\newtheorem{question}{\textbf{RQ}}
\usepackage{listings}
\usepackage{xcolor}
\usepackage[ruled,vlined]{algorithm2e}

\usepackage{multirow}
\usepackage{subfig}
\usepackage{amssymb}
\usepackage{graphicx}
\usepackage{booktabs}
\usepackage{tcolorbox}
\usepackage{threeparttable}
\usepackage[font=footnotesize,labelfont=bf]{caption}

\usepackage{tikz}
\newcommand*\emptycirc[1][1ex]{\tikz\draw (0,0) circle (#1);} 
\newcommand*\halfcirc[1][1ex]{%
	\begin{tikzpicture}
	\draw[fill] (0,0)-- (90:#1) arc (90:270:#1) -- cycle ;
	\draw (0,0) circle (#1);
	\end{tikzpicture}}
\newcommand*\fullcirc[1][1ex]{\tikz\fill (0,0) circle (#1);}

\newtheorem{myDef}{Definition}
\newcommand{\tabincell}[2]{\begin{tabular}{@{}#1@{}}#2\end{tabular}}
\usepackage{makecell}
\usepackage{color}

\def\BibTeX{{\rm B\kern-.05em{\sc i\kern-.025em b}\kern-.08em
    T\kern-.1667em\lower.7ex\hbox{E}\kern-.125emX}}
\begin{document}

\title{Interpretation-enabled Software Reuse Detection Based on a Multi-Level Birthmark Model}

\author{ Xi~Xu \IEEEauthorrefmark{1}\IEEEauthorrefmark{2}, Qinghua~Zheng \IEEEauthorrefmark{1}\IEEEauthorrefmark{2}, Zheng~Yan  \IEEEauthorrefmark{4}\IEEEauthorrefmark{5}, Ming~Fan \IEEEauthorrefmark{1}\IEEEauthorrefmark{3},   
Ang~Jia \IEEEauthorrefmark{1}\IEEEauthorrefmark{3}, 
and Ting~Liu \IEEEauthorrefmark{1}\IEEEauthorrefmark{3}
\thanks{Corresponding author: Zheng~Yan.}
\IEEEauthorblockA{\\\IEEEauthorrefmark{1}Key Laboratory of Intelligent Networks and Network Security, Ministry of Education, China}
\IEEEauthorblockA{\IEEEauthorrefmark{2}School of Computer Science and Technology, Xi'an Jiaotong University, China}
\IEEEauthorblockA{\IEEEauthorrefmark{3}School of Cyber Science and Engineering,  Xi'an Jiaotong University, China}
\IEEEauthorblockA{ \IEEEauthorrefmark{4}State Key Lab on Integrated Services Networks, School of Cyber Engineering, Xidian University, China}
\IEEEauthorblockA{\IEEEauthorrefmark{5}Department of Communications and Networking, Aalto University, Finland}

\IEEEauthorblockA{xx19960325@stu.xjtu.edu.cn; qhzheng@xjtu.edu.cn; zyan@xidian.edu.cn;  \\ mingfan@mail.xjtu.edu.cn; jiaang@stu.xjtu.edu.cn;  tingliu@mail.xjtu.edu.cn}
}

\maketitle

\begin{abstract}
Software reuse, especially partial reuse, poses legal and security threats to software development. Since its source codes are usually unavailable, software reuse is hard to be detected with interpretation. On the other hand, current approaches suffer from poor detection accuracy and efficiency, far from satisfying practical demands. To tackle these problems, in this paper, we propose \textit{ISRD}, an interpretation-enabled software reuse detection approach based on a multi-level birthmark model that contains function level, basic block level, and instruction level. To overcome obfuscation caused by cross-compilation, we represent function semantics with Minimum Branch Path (MBP) and perform normalization to extract core semantics of instructions. For efficiently detecting reused functions, a process for ``intent search based on anchor recognition'' is designed to speed up reuse detection. It uses strict instruction match and identical library call invocation check to find anchor functions (in short anchors) and then traverses neighbors of the anchors to explore potentially matched function pairs. Extensive experiments based on two real-world binary datasets reveal that \textit{ISRD} is interpretable, effective, and efficient, which achieves $97.2\%$ precision and $94.8\%$ recall. Moreover, it is resilient to cross-compilation, outperforming state-of-the-art approaches.

\end{abstract}


\section{Introduction}

Along with the growing popularity of open-source software, software reuse becomes a common phenomenon. However, extensive reuse of existing codes leads to numerous license violation issues~\cite{hemel2011finding, duan2017identifying}. For example, Cisco and VMWare were exposed to significant legal issues because they did not adhere to the licensing terms of Linux kernel~\cite{Cisco, VMware}. What is more, security issues could be raised due to careless software reuse~\cite{li2016vulpecker}.

The goal of software reuse detection is to determine whether a candidate program contains similar codes already used in a target program. There are many approaches~\cite{kamiya2013agec, su2016code} proposed in the literature. According to the analysis objects, these approaches can be divided into two groups: source code reuse detection and binary code reuse detection. The first calculates code similarity by abstracting source codes into a set of characteristics, such as string~\cite{ducasse1999a, baker1995on},  token~\cite{brixtel2010language-independent,cosma2012an,wise1996yap3,whale1990identification}, Abstract Syntax Tree (AST)~\cite{zhang2012ast-based,kikuchi2014a, jiang2007deckard, koschke2006clone} and Program Dependency Graph (PDG)~\cite{krinke2001identifying,komondoor2001using,gabel2008scalable}. Since the source code of a candidate program is typically unavailable in reality, binary code reuse detection is used widely for software plagiarism detection~\cite{luo2014semantics, tian2013dkisb,wang2009detecting}, malware
detection~\cite{kruegel2005polymorphic, bruschi2006detecting, ming2015memoized}, patch analysis~\cite{gao2008binhunt,hu2016cross-architecture,xu2017spain}, and so on.

However, the existing binary code reuse detection approaches suffer from three limitations, as described below.

\emph{Poor interpretability}: The detection results of the existing approaches~\cite{luo2014semantics, kruegel2005polymorphic, gao2008binhunt} lack of the ability to provide detailed evidence to support reuse detection results because they usually only report their results in form of similarity scores ranging from 0 to 1. The ability to comprehensively interpreting the detection results is extremely important since it can provide the details or reasons to make the detection results acceptable or easy to be understood.

\emph{Poor accuracy}: Most approaches~\cite{tamada2004design,david2014tracelet,lim2009method, lim2009a,bindiff} that rely on structural and syntax information fail to deal with the differences caused by variations in compilation. This is because different compilations of a source program naturally produce different structures and syntax in its binary codes, forming obfuscation. For example, the approaches proposed in \cite{david2017similarity, david2016statistical} that operate at the boundaries of a basic block might fail to deal with basic block splitting or merging.

\emph{Poor efficiency}: Several works~\cite{bindiff,david2017similarity} use a brute force method to identify reused functions, which is prohibitively expensive since it measures the similarities of all function pairs between a target program and a candidate program. Such approaches lead to poor efficiency if the numbers of functions in both programs are big. Therefore, these approaches are neither effective in case that a reused part only makes up a small percentage of the candidate program, nor efficient if an excessive number of function pairs need to be compared.  

To overcome the above limitations, we propose \textit{ISRD}, a novel Interpretation-enabled Software Reuse Detection approach based on a multi-level birthmark model, which holds the following salient advantages:

\emph{Interpretable detection results.} \textit{ISRD} is capable of capturing program semantics from coarse granularity to fine granularity and uniquely identifying a program with a multi-level birthmark model that contains function level, basic block level, and instruction level. Specifically, at the function level, a Function Call Graph (FCG) is constructed to profile program behavior. The FCG of a program is a directed graph, which consists of a set of nodes representing functions and a set of edges representing caller and callee relationships among functions~\cite{shang2010detecting}. Then, basic block chains and normalized instructions are extracted to demonstrate the semantics of the function at the basic block level and the instruction level, respectively. The similar parts between the birthmark of a target program and that of a candidate program in the above three levels can reconstruct a reuse scene to interpret and justify detection results. Obviously, this kind of demonstration can assist easy understanding and trust on a detection result, unlike a simple value reuse indicator.

\emph{High accuracy.} To achieve high accuracy of reuse detection, we perform normalization at different levels of our birthmark model. Specifically, at the basic block level, we first transform each function into a set of Minimum Branch Paths (MBPs), which are length variant partial execution paths starting from an initial node or a branch node and ending at a terminal node or an adjacent branch node. Then, to mitigate huge differences in instructions caused by cross-compilation, we only consider the key instructions to represent their core semantics. Furthermore, we lift low-level assembly instructions up to high-level operations and remove operands from them to address syntax differences of instructions to complete the process of instruction normalization.

%we normalize them as below. First, we only consider the key instructions in order to retain their core semantics. Then, we lift low-level assembly instructions up to high-level operations and remove operands from them in order to address syntax differences of instructions.

\emph{High efficiency.} To speed up the similarity calculation among thousands of function pairs, we propose a process for ``intent search based on anchor recognition'' to first recognize anchor functions (in short anchors) and then perform intent search originated from the anchors to discover all potentially matched function pairs. In this way, we significantly reduce the number of comparisons before similarity calculation and meanwhile ensure comparison quality. Concretely, the process leverages on strict instruction match and identical library call invocation check to search for matched function pairs as anchors by considering both developer-defined functions and library functions. Through intent search originated from the anchors, it further explores neighbors of the anchors to find new function pairs with high similarity scores.

Moreover, since there is currently no dataset that can be directly used for partial reuse detection tests, we construct a dataset that contains 24 real-world software projects and manually label a total of 74 partial reuses. The dataset has been published on website~\cite{Dataset}. By evaluating \textit{ISRD} based on our constructed dataset and a widely used dataset, we demonstrate that \textit{ISRD} exhibits impressive software reuse detection performance. In summary, the major contributions of this paper include:
\renewcommand\theenumi{\roman{enumi}}
\renewcommand\labelenumi{(\theenumi)}
\begin{enumerate}

	\item We propose a novel fine-granular multi-level birthmark model to uniquely represent program semantics and enable interpretability of software reuse detection result. 
	
	\item We perform normalization at both the basic block level and the instruction level to resist semantics-preserving obfuscation for accurate software reuse detection.
	
	\item We design a process to recognize anchors and conduct intent search originated from the anchors, which can greatly reduce the number of comparisons, thus significantly accelerate detection speed.
	
	\item We implement \textit{ISRD} and evaluate its performance with extensive experiments. The results reveal that \textit{ISRD} is interpretable, can effectively and efficiently detect partial reuse. It is also resilient to cross-compilation, outperforming the state-of-the-art approaches.
\end{enumerate}

\section{Related Work}
	\label{Sec_RelatedWork}
%According to the form of the target program and the candidate program, existing software reuse detection approaches can be divided into two main categories: source  code reuse detection and binary code reuse detection. We briefly review the two types of existing approaches in this section and compare them with \textit{ISRD} in a qualitative way.
%Since \textit{ISRD} is an approach to binary code reuse detection, 
%We briefly review the existing related work and compare them with \textit{ISRD} in a qualitative way.

%\subsubsection{Source Code Reuse Detection}
%\textit{Moss}~\cite{Moss} is an automatic system for detecting software reuse. It applies a local document fingerprinting algorithm to capture an essential property of any fingerprinting technique guaranteed to detect reuse. \textit{JPlag}~\cite{Prechelt2000JPlag} is a system that finds pairs of similar programs among a given set of programs by parsing the programs and converting them into token strings. Zhang et al.~\cite{zhang2012ast} proposed \textit{AST}, which studies the code reuse relationship by converting the code to AST. Liu et al.~\cite{liu2006gplag} proposed \textit{GPLAG}, which detects reuse by mining program dependence graphs (PDGs). The similarity between PDGs is calculated by using graph isomorphism algorithms.

% \subsubsection{Binary Code Reuse Detection}
According to analysis techniques, existing binary code reuse detection approaches can be divided into two main categories: static analysis and dynamic analysis.
 
\emph{Static Analysis.} Static analysis is applied on binaries without running programs. Bindiff~\cite{bindiff} was proposed to use the structural similarity of CFG to compare binary codes. Luo et al.~\cite{luo2014semantics} proposed \textit{CoP}, a binary-oriented, obfuscation-resilient method for code reuse detection, which combines rigorous program semantics with the longest common subsequence based fuzzy matching. David et al.~\cite{david2016statistical} proposed a new approach \textit{Esh} of calculating binaries' similarity, which firstly decomposes binary codes into small comparable fragments, then defines semantic similarity between fragments, and further uses statistical reasoning to lift fragment similarity into the similarity between procedures. Chandramohan et al.~\cite{chandramohan2016bingo} proposed \textit{Bingo}, which captures complete function semantics by inlining the relevant library and developer-defined functions and models binary functions in a program structure agnostic fashion by using length variant partial traces. Feng et al.~\cite{feng2016scalable} proposed \textit{Genius} to search vulnerabilities in massive IoT ecosystems by converting Attributed Control Flow Graph (ACFG) into high-level numeric feature vectors. Different from \textit{Genius} that embeds an ACFG by taking a codebook-based approach, Xu et al.~\cite{xu2017neural} proposed \textit{Gemini} to take a neural network-based approach to transform the ACFGs into embeddings of binary functions for similarity detection. Liu et al.~\cite{liu2018alphadiff} proposed a solution named \textit{$\alpha$Diff}, which employs three semantic features: intra-function feature, inter-function feature and inter-module feature, to address cross-version binary code similarity detection challenges.

\begin{table}[t]
    \caption{Comparison with Existing Approaches}
    \scriptsize
    \label{Tab-Sec2-RelatedWork}
    \centering
    
    \begin{threeparttable}
    \scalebox{1}{
      \begin{tabular}{p{1.0cm}<{\centering}  p{0.8cm}<{\centering} p{0.8cm}<{\centering}p{0.8cm}<{\centering}p{0.8cm}<{\centering}p{0.8cm}<{\centering}}
      
        	\toprule[1.5pt]

       Approach     
			& Type \tnote{1}  &    Granu. \tnote{2}    &  Inter.\tnote{3}    &   Effec. \tnote{3}  & Effic. \tnote{3}\\  \midrule
\textit{Bindiff}~\cite{bindiff}&St  & B &  \emptycirc &   \emptycirc  & \emptycirc \\
\textit{Cop}~\cite{luo2014semantics}&St & B &  \emptycirc &  \fullcirc  & \emptycirc \\
\textit{Esh}~\cite{david2016statistical}&St & I & \emptycirc & \halfcirc &  \emptycirc\\
\textit{Bingo}~\cite{chandramohan2016bingo} &St & B &  \emptycirc &  \halfcirc  & \halfcirc\\
\textit{Genius}~\cite{feng2016scalable} &St & B &  \emptycirc  &  \halfcirc  & \fullcirc \\
\textit{Gemini}~\cite{xu2017neural}&St & B &  \emptycirc  &  \halfcirc  & \fullcirc\\
\textit{$\alpha$Diff}~\cite{liu2018alphadiff} &St & N.A. &  \emptycirc &  \halfcirc  &\fullcirc \\
\textit{DYKIS}~\cite{tian2015software} &Dy & - &  \emptycirc &   \fullcirc & \emptycirc  \\
\textit{TOB}~\cite{tian2017reviving} &Dy & -&  \emptycirc &  \fullcirc &  \emptycirc \\
\textit{LoPD}~\cite{2016Deviation} &Dy & -&  \emptycirc &  \fullcirc  &  \emptycirc  \\ \midrule
\textit{ISRD} &St& I, B, F & \fullcirc &  \fullcirc  &  \fullcirc \\
        
        \bottomrule[1.5pt]
      \end{tabular}
      }
   
    \begin{tablenotes}
    \footnotesize
    \item[1] St: Static; Dy: Dynamic.
    \item[2] Granu.: Granularity; I: Instruction Level; B: Basic Block Level; F: Function Level.
    \item[3]  Inter.: Interpretability; Effec.: Effectiveness; Effic.: Efficiency; \fullcirc, \halfcirc\,and\,\emptycirc\,respectively represent satisfying the criteria, partially satisfying the criteria, and  not satisfying the criteria.
     
    \end{tablenotes}

    \end{threeparttable}
    \vspace{-15pt}
  \end{table}
  
\emph{Dynamic Analysis.} A dynamic approach for binary code reuse detection is performed during code execution by running programs. Tian et al.~\cite{tian2015software,tian2013dkisb} proposed \textit{DYKIS} to uniquely identify a program for detecting similar programs. Tian et al.~\cite{tian2017reviving} presented a framework called \textit{Thread Oblivious dynamic Birthmark (TOB)} that revives existing techniques to detect reuse of multi-thread programs. Ming et al.~\cite{2016Deviation} proposed a logic-based approach \textit{LoPD} by leveraging dynamic symbolic execution and theorem proving techniques to capture dissimilarities between two programs in order to rule out semantically different programs. 

\begin{figure*}[t]
	\centering
	\includegraphics[width=0.85\textwidth]{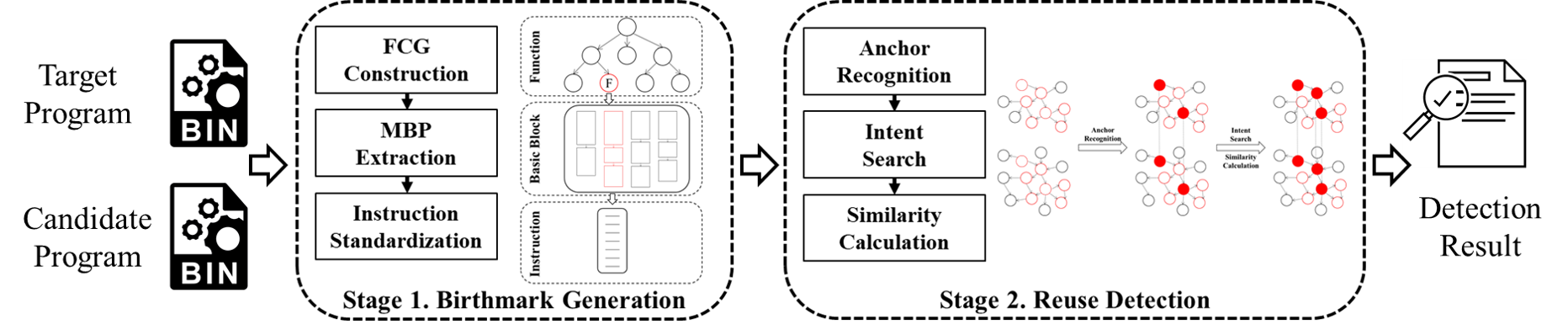}
	\centering
	\caption{\textbf{\textit{ISRD}} Overview}
	\label{Fig-Sec3-Overview}
	\vspace{-15pt}
\end{figure*}

For better understanding the differences of the above approaches from \textit{ISRD}, we compare them in Table \ref{Tab-Sec2-RelatedWork} in terms of  analysis type, granularity of applied birthmark,  interpretability, effectiveness for resisting obfuscation caused by cross-compilation, and efficiency. From Table~\ref{Tab-Sec2-RelatedWork}, we observe that the existing approaches cannot interpret detection results, although they can resist obfuscation caused by cross-compilation to some extent. \textit{Genius}~\cite{feng2016scalable}, \textit{Gemini}~\cite{xu2017neural} and \textit{$\alpha$Diff}~\cite{liu2018alphadiff} use deep learning~\cite{lecun2015deep} to detect reuse and achieve high efficiency, but they cannot interpret detection results. None existing approach can comprehensively overcome the aforementioned limitations except for \textit{ISRD}, which shows the novelty and advantages of \textit{ISRD}.

\section{\textit{ISRD} Design}
\label{Sec_Approach}

In this section, we introduce the technical details of \textit{ISRD}, which contains two main stages as illustrated in Fig.\ref{Fig-Sec3-Overview}.

\textit{Birthmark Generation.} This stage constructs the birthmarks of the target program and the candidate program. The birthmark relates to three levels, i.e., function level, basic block level, and instruction level. At the function level, an FCG is constructed to depict the program semantics. At the basic block level, the MBPs of a given function are extracted to profile its behaviors. At the instruction level, normalization is performed to capture instruction core semantics.

\textit{Reuse Detection.} Since the direct comparison of all function pairs between two programs is a time-consuming job, a process for ``intent search based on anchor recognition'' is proposed to recognize anchors and conduct intent search originated from the anchors to significantly accelerate function pair matching. Specifically, in \textit{Anchor Recognition}, strict instruction match and identical library call invocation check are used to find the anchors. Then, in \textit{Intent Search}, we originate from the anchor pairs to explore potentially matched function pairs based on their function call relationships.

\subsection{Birthmark Generation}
A birthmark is a set of characteristics extracted from a program that reflects its semantic behaviors, which can be used to uniquely identify a program and is resilient to semantics-preserving code transformations. 

In the literature, there are many proposed birthmarks with different granularities to represent a program. But in practice, applying coarse granularity may restrict program similarity catching in a precise way. For example, a program-level birthmark is unable to detect partial reuse, since a candidate program often reuses only a part of a target program. Meanwhile, the birthmark similarity at a fine granularity level cannot be used to infer the similarity at a coarse granularity level. For example, given the similarity at instructions, we cannot conclude that the functions of two programs are similar. 

To address this problem, we propose a multi-level birthmark model to characterize a program by involving three-level granularities to form a hierarchical birthmark, from the function level, to the basic block level, and finally the instruction level. 

On the top of the hierarchical birthmark, to depict the program semantics from the point of a macroscopic view, we first construct the FCG of a program to describe its behavior. Then, we turn our attention to the individual functions in the program by distilling the function semantics with MBP representation. Finally, to further capture the instruction semantics, we perform normalization on the finest-granularity instructions. Consequently, a program can be identified by a three-level birthmark that contains function birthmark, basic block birthmark, and instruction birthmark. Next, We discuss the details of birthmark at each level.   

\subsubsection{FCG Construction}
To capture program semantics at the function level, we construct FCG for both the target program and the candidate program. The FCG captures the functionality and objective of a program semantically from its structural information to profile program behaviors~\cite{fan2019graph, fan2018android, fan2017dapasa}. 

In order to construct FCG of a given program, we first identify the invocation statements (i.e., ``call'') from its assembly code to extract callers and callees. Then, the callers and the callees are added into a graph as nodes. In addition, if a function call relation exists between a caller and a callee, an edge is inserted between them in the graph. 

\subsubsection{MBP Extraction}
To characterize function semantics, Control Flow Graph~(CFG) is applied, which contains detailed information of the basic blocks in a function. In a CFG, each node represents a basic block that is a straight-line piece of code without any branch, and each edge represents the control flow relationship among blocks. A CFG is defined as follows.

\begin{myDef}
	\label{Def-CFG}
	\textbf{Control Flow Graph~(CFG)}: A CFG is a directed graph $G=(V,E)$. 	
	\begin{itemize}
		\item{$V=\{v_i|1\le i \le n\} $ denotes the set of basic blocks of a function, where $v_i \in V $ is the $i$th block.}
		\item{$E\subseteq V\times V $ denotes the set of control flows, where $ (v_i,v_j)\in E $ indicates that a control flow is from $v_i$ to $v_j$.}\hfill$\blacksquare$	
	\end{itemize}
\end{myDef}

However, existing approaches ~\cite{david2016statistical,david2017similarity} that operated at the boundaries of basic blocks are vulnerable to block splitting or merging when different compilation processes are applied. Our solution to this problem is inspired by \textit{TRACY}~\cite{david2014tracelet}, which uses \textit{tracelets} -- partial traces of an execution to compare function similarity. Concretely, we propose MBPs that are partial straight-line execution paths between branching nodes in a CFG. A branching node is a node that has more than one successor node.

To extract MBPs, we firstly pre-process the CFG by grouping basic blocks into a number of straight-line paths (i.e., replacing edges that connect two blocks with a single out-edge and a single in-edge, respectively). This kind of grouping does not affect the function semantics and is only for the purpose of simplifying the generated MBPs, which is defined as below. 

\vspace{5 pt}
\begin{myDef}
	\label{Def-MBP}
	\textbf{Minimum Branch Path (MBP)}: Let a path extracted from a $CFG$ be 
	denoted as a node sequence $p=\langle v_{p_1}, \dots, v_{p_n} \rangle$. 
	A path $p$ is a MBP if the following conditions are satisfied:  	
	\begin{itemize}
		\item{$v_{p_1}$ is an initial node, which has no predecessor or is a branching node.}
		\item{$v_{p_n}$ is a terminal node,  which has no successor or is a branching node.}
		\item{No other nodes in $p$ are initial or terminal nodes.}\hfill$\blacksquare$		
		% Other nodes in $p$ denote the nodes which have and only have one successor node and preccessor node. .}
	\end{itemize}
	%\vspace{5pt}
	%	\vspace{-8pt}
\end{myDef}

 Unlike the fix-lengthed \textit{tracelet} proposed in \cite{david2014tracelet}, MBP is variable in length according to the structure of CFG. It has a number of characteristics making it suitable for representing function semantics:

\begin{itemize}
    \item {\textit{Semantics Exhibition}: The combination of basic blocks and control flow in MBP can represent the execution of a function and capture its semantics.}
    \item {\textit{Effectiveness and Efficiency}: Trying to gather and analyze all paths in a CFG is clearly infeasible. MBP effectively cuts down the size and the number of paths of CFG by only considering sub-paths between branching nodes. Thus, it can help in speeding up function semantics matching.}
    \item {\textit{Structural Variation Resilience}: The absence of branches in MBP implies that it would be less vulnerable than CFG to structural changes caused by block splitting and merging.}
    \item {\textit{Resilience to Jump Instruction Variations}: Jump instructions are sensitive to obfuscation. MBP is by nature free from jump instructions due to branch omission. }
\end{itemize}

Algorithm~\ref{Alg-MBP} shows the steps of extracting MBPs from a given CFG. The output $P$ denotes the set of extracted MBPs. The  functions \textit{In} and \textit{Out} return the in-degree and out-degree of a given node. Specifically, $P$ is initialized as an empty set. Then, for every node $v_i$ in $V$, if $v_i$ has no direct predecessor nodes or more than one direct successor node, the function \textit{EXTRACT} is invoked to extract MBPs from $v_i$. Finally, the extracted MBPs are added to $P$.   

\begin{algorithm}[t]
\footnotesize
	\DontPrintSemicolon
	\caption{MBP Extration from CFG}%算法名字
	\label{Alg-MBP}
	\LinesNumbered %要求显示行号
	\SetKwFunction {EXTRACT}{EXTRACT} 
	\SetKwProg{Fn}{Function}{:}{}	
	\KwIn{$G =(V,E)$\quad \textcolor[rgb]{0.5,0.5,0.5} {// $G$ is a CFG.} }%输入参数
	\KwOut{$P$ \quad\textcolor[rgb]{0.5,0.5,0.5} {// $P$ is the set of all MBPs.}}%输出
	$ P=\emptyset $\; %\;用于换行
	\For{each $v_i\in V$}{
		\If{\textbf{In}$(v_i)=0$ or \textbf{Out}$(v_i)>1$}{
			$\Call{EXTRACT}{v_i}\rightarrow P$\;					
		}
	}
	\KwRet {$P$}\;
	
	\SetKwProg{Fn}{Function}{:}{}
	\Fn{\EXTRACT{$v_i$}}{
		$P_i=\emptyset $\;
		\For {each $(v_i,v_j) \in E$} 
		{
			$v_i,v_j \rightarrow p$\;
			\While{\textbf{Out}$(v_j)=1$}
			{
				$v_{k,(v_j,v_k) \in E} \rightarrow p$	\;
				$v_j=v_k$
			}
			$p \rightarrow P_i $\;
		}
		\KwRet $P_i$\;
	}
\vspace{-3pt}
\end{algorithm}

\begin{figure}[t]

	\centering
	\includegraphics[width=0.45\textwidth]{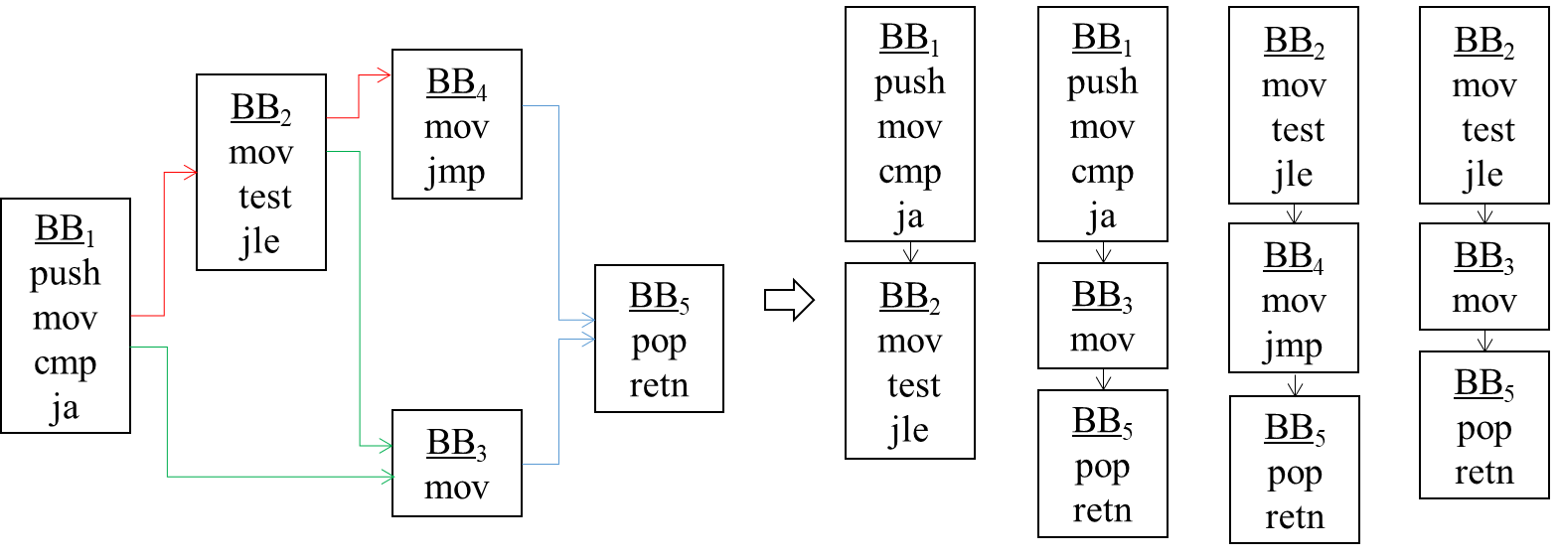}
	\centering
	\caption{A sample CFG and its extracted MBPs}
	\label{Fig-Sec3-MBP}
	\vspace{-15pt}
\end{figure}

Fig.~\ref{Fig-Sec3-MBP} depicts an example of CFG and its extracted MBPs. We observe that the original basic blocks in a control-flow are grouped as execution flows, which are already determined. There are two basic blocks $BB_1$, $BB_2$ that have more than one direct successor basic blocks. Therefore, the extracted MBPs from the CFG are: $(BB_1,BB_2)$, $(BB_1, BB_3, BB_5), (BB_2, BB_4, BB_5).(BB_2, BB_3, BB_5)$.

\subsubsection{Instruction Normalization}
Syntax is the most direct birthmark to represent instruction semantics. However, different compilation processes may cause significant differences in the assemblies~\cite{david2014tracelet,tian2015software}. To retain the core semantics of instructions and be resilient to obfuscation introduced by cross-compilation, normalization over instructions is applied.% in \textit{ISRD}.

\emph{Key Instruction Extraction.} First, equally treating all kinds of instructions  may be defeated by compilation variations. That is because some kinds of instructions are not closely related to semantics and are easily changed across compilation. To solve this problem, we only consider key instructions. Ideally, the key instructions should constitute a small portion of a whole execution sequence and must be relatively unique. Through observation, we find that there exist a large number of data-transfer instructions, such as \textit{push} and \textit{pop}, especially \textit{mov}, in almost all MBPs. These instructions can be discarded because they usually facilitate computations rather than belong to a part of MBP logic. Furthermore, they are easily added and deleted compared with other instructions.

Actually, deleting all data-transfer instructions is probably the simplest approach to solve the problem caused by cross-compilation, but it might lead to the loss of data-transfer semantics. Thus, we only keep the first data-transfer instruction when multiple data-transfer instructions appear continuously. 

\emph{Instruction Lifting.}  As we discussed earlier, the same operation can be expressed in different instructions. To achieve semantics equivalence on instructions, we lift the instructions into high-level operations. For example, two instructions  \textit{inc} and \textit{add} can be mapped to one addition operation. 

\emph{Operand Removing.}  The operands in the instructions are easily changed in the compilation. Even compiling the same source code with the same compilation settings, the operands can be different due to their differences in memory layout. Therefore, we strip the operands from the instructions.

At this point, by using the proposed multi-level birthmark model, we construct the three-level birthmarks for both target and candidate programs, which give a detailed description of the program semantics from coarse granularity to fine granularity.

\subsection{Reuse Detection}
\label{Subsec_ReuseDetection}
After \textit{Birthmark Generation}, given the target program and the candidate program, we can capture their similarity relationship based on their birthmarks for reuse detection.

To trade off between coarse granularity and fine granularity, it is reasonable to take the function as a comparison unit. In order to compare functions, we first perform comparison at the instruction level, then combine the comparison results to compute the similarity at the basic block level, and then the function similarity is determined. Finally, the comparison results between functions are aggregated to capture the similarity between the target program and the candidate program. In addition, the matching relationship at the three levels is presented to interpret the detection results. 

However, matching target functions with a large number of candidate functions remains a major bottleneck. The reason is that the scale of function pair-wise comparison increases exponentially with the number of functions.
 
To tackle this problem, an efficient matching process that significantly accelerates the search for matched function pairs between the target and the candidate programs is proposed. The process contains two main steps, i.e., \textit{Anchor Recognition} and \textit{Intent Search}, to predict which function pairs are likely to match, thereby making the matching process much more efficient by avoiding unnecessary matching. The process first performs strict instruction match and identical library call invocation check to identify matched anchor function pairs, then it explores neighbors of the matched anchors to find new function pairs with a high similarity score under the guidance of function level birthmark. By doing this, we can effectively reduce the number of comparisons before program similarity score calculation to make reuse detection execute swiftly.

\subsubsection{Anchor Recognition}
This step attempts to find matched anchor function pairs, which can efficiently instruct later reused function matching. Specifically, we jointly use two ways to recognize the anchors among a huge number of functions in the candidate program.

\textit{\textbf{Way 1}}: Given a function in the target program, we search the functions in the candidate program that meet strict instruction matching. Here, the strict instruction matching indicates that both the numbers of instructions and their sequences in two functions are exactly the same.

\textit{\textbf{Way 2}}: Considering that library call invocations provide an important partial semantics of a function~\cite{2016Using}, we also look for identical library call invocations that appeared in both the candidate program and the target program. 

Note that \textit{\textbf{Way 1}} and \textit{\textbf{Way 2}} operate on the developer-defined functions and library functions, respectively.

\begin{algorithm}[t]
  \footnotesize
	%	\DontPrintSemicolon
	\DontPrintSemicolon
	%\SetAlgoLined
	\caption{Anchor Recognition}%算法名字
	\label{Alg-Anchor}
	\LinesNumbered %要求显示行号
%	\SetKwFunction {getFuncIns}{getFuncIns} 
%	\SetKwFunction {getLibFunc}{getLibFunc} 
%	\SetKwProg{Fn}{Function}{:}{}	
	\KwIn{ 
		
		$D_{T}, L_{T}$
		\quad \textcolor[rgb]{0.5,0.5,0.5} {// $D_{T},L_{T} $ denotes the developer-defined functions and library functions in the target program.} 
		
		$D_{C}, L_{C}$
		\quad \textcolor[rgb]{0.5,0.5,0.5} {// $D_{C}, L_{C}$ denotes the developer-defined functions and library functions in the candidate program.} 
	}
	
	\KwOut{
	
	$Anchor$ \quad\textcolor[rgb]{0.5,0.5,0.5} {//$Anchor$ denotes the set of matched anchor function pairs}}
	
	$ Anchor = \{\} $\; %\;用于换行
	\ForEach{$d_T \in D_T$}
	{
	$ I_{d_T} = GetFuncIns(d_T)$\;
			\ForEach {$d_C  \in D_C$}
		{
		$ I_{d_C} = GetFuncIns(d_C)$\;
		
		$sim_{ins}  = SimIns(I_{d_T},I_{d_C}) $\;
			\If{$sim_{ins}==1$}
			{	
				$ Anchor  \leftarrow (d_T,d_C) $\;
			}
		}

	}
    \ForEach{$l_T \in L_{T}$}
	{
	    \ForEach{$l_C \in L_{C}$}
	    {
		\If{$ l_C == l_T$}
		{
			$Anchor  \leftarrow (l_T,l_C)$ \;	
		}
		}
	}
\vspace{-3pt}
\end{algorithm}

Algorithm~\ref{Alg-Anchor} shows the process of \textit{\textbf{Anchor Recognition}}. The inputs $D_{T}$ and $L_{T}$ denote the developer-defined functions and library functions in the target program, respectively. $D_{C}$ and $L_{C}$ denote the developer-defined functions and library functions in the candidate program, respectively. The output $Anchor$ denotes a set of matched anchor function pairs. The function \textit{GETFUNCINS} takes a function as input and returns its all instructions. At line 6, we use Jaccard distance to measure the similarity between the function pairs in terms of \textit{\textbf{Way 1}}, i.e., their similarity in instructions. After that, we obtain the identical library function invocations in both the target program and the candidate program (lines 9-12).

\subsubsection{Intent Search}
In the intent search, we originate from the matched anchors to discover new matched function pairs. \textit{ISRD} loops over each recognized anchor and explores its direct neighbors to find new matched function pairs that have a high similarity score. The valid correspondences found in the previous step are propagated to their neighbors. And the new identified matched function pairs are added as new anchors.

Given a matched anchor function pair $A$ and $a$, the basic idea of intent search is that the neighboring functions of $A$ that are connected to $A$ with an edge in the function level birthmark usually correspond to those of $a$.

In order to search for new matched function pairs, we use the function invocation relationship to guide our processing. We consider the cues got from the current state of the matched function pairs and the function call relationship in FCG. Instead of exploring completely at random, we set a priority to guide the search and improve its efficiency by increasing the probability of finding matched function pairs. We give a high priority to the functions that have the most number of matched neighbors. Subsequently, the process should firstly operate on the high priority functions in the candidate program. With this process, all matched function pairs between the target program and the candidate program can be identified.

Algorithm~\ref{Alg-Intent} shows the process of \textit{\textbf{Intent Search}}. The input $Anchor$ denotes the set of matched anchor function pairs; $f_{T_{B}}$ denotes the function that has the highest priority in the target program; $FCG_T$ and $FCG_C$  denote the function level birthmarks of the target program and the candidate program, respectively. The output $FF$ denotes a set of potentially matched function pairs of $f_{T_{B}}$.

Firstly, for each precursor node $f_{T_{A}}$ of $f_{T_{B}}$ in $FCG_T$, if $( f_{T_{A}},  f_{C_{a}}) $ is in $Anchor$, $(f_{T_{B}}, f_{C_{b}}) $ is added in $Pre$, where $f_{C_{b}}$ is the successor node of  $ f_{C_{a}}$~(lines 2-3).

Then for each  successor node $f_{T_{D}}$ of $f_{T_{B}}$ in $FCG_T$, if $( f_{T_{D}},  f_{C_{d}}) $ is in $Anchor$, $(f_{T_{B}}, f_{C_{b}}) $ is added in $Suc$, where $f_{C_{b}}$ is the precursor node of  $ f_{C_{d}}$~(lines 4-5).

Finally, if $Pre$ and $Suc$ are not empty, the intersection of them is added into $FF$, otherwise their union is added into $FF$~(lines 6-9). 

\begin{algorithm}[t]
\footnotesize
	%	\DontPrintSemicolon
	\DontPrintSemicolon
	%\SetAlgoLined
	\caption{Intent Search}%算法名字
	\label{Alg-Intent}
	\LinesNumbered %要求显示行号
	\SetKwFunction {getFunc}{getFunc} 
	\SetKwProg{Fn}{Function}{:}{}	
	\KwIn{ 
	
		$Anchor$
		\quad \textcolor[rgb]{0.5,0.5,0.5} {// $Anchor$ denotes the matched anchor function pair set.} 
		
		$f_{T_{B}}$
		\quad \textcolor[rgb]{0.5,0.5,0.5} {// $f_{T_{B}}$ denotes the function which has the highest priority in the target program.}
		
		$FCG_T, FCG_C$ \quad \textcolor[rgb]{0.5,0.5,0.5} {// $FCG_T, FCG_C$ denote the function level birthmark of the target program and the candidate program.}
 }
	
	\KwOut{
	
	$FF$ \quad\textcolor[rgb]{0.5,0.5,0.5} {//$FF$ denotes a set of  potential matched function pair of $f_{T_{B}}$. }}
	
	$ FF, Pre, Suc = \{\} $\; 
	\ForEach{ $( f_{T_{A}}, f_{T_{B}})_{( f_{T_{A}},  f_{C_{a}}) \in Anchor}\in FCG_T $}
	{
	 $Pre \leftarrow (f_{T_{B}}, f_{C_{b}})_{( f_{C_{a}}, f_{C_{b}}) \in FCG_C } $
	}
	\ForEach{$ (f_{T_{B}},  f_{T_{D}})_{( f_{T_{D}},  f_{C_{d}}) \in Anchor} \in FCG_T $
	}{
	 $Suc \leftarrow (f_{T_{B}}, f_{C_{b}})_{( f_{C_{b}}, f_{C_{d}}) \in FCG_C }$
	}
	\If{$len(Pre) \textgreater 0 $ and $len(Suc) \textgreater 0 $ }
	{ $FF \leftarrow Pre \cap Suc$}
	\Else{  $FF \leftarrow Pre \cup  Suc$} 

\vspace{-3pt}
\end{algorithm}

\subsubsection{Similarity Calculation}
After identifying the potentially matched function pairs, we are able to measure the similarities of two functions by calculating the similarity scores based on their basic block level birthmarks - MBP sets. To compute a similarity score for each pair of
MBP sets, we first show how to compute the similarity score for a pair of MBPs using Equation~(\ref{simmbp}).
\begin{equation}
\small
\label{simmbp}
\begin{aligned}
sim(mbp_1, mbp_2) =\frac{ 2 \times lcs(mbp_1, mbp_2)}{|mbp_1|+| mbp_2|}
\end{aligned}
\end{equation}

We compute the Longest Common Subsequence (LCS) of two MBPs by utilizing the LCS dynamic programming algorithm, denoted as $lcs(mbp_1, mbp_2)$. Then, the similarity score between two MBPs is calculated by taking the length of LCS divided by the average length of two MBPs. By trying each pair of MBPs, we use the collected individual similarity scores to calculate the similarity of two MBP sets ($MBP_1$, $MBP_2$) according to the following Equation ~(\ref{simMBP}-\ref{maxscore}).
\begin{equation}
\small
\label{simMBP}
\begin{aligned}
sim(MBP_1, MBP_2) =  \sum_{mpb_1 \in MBP_1}\frac{|mpb_1| MaxScore }{\sum_{mpb_1 \in MBP_1}|mpb_1|}
\end{aligned}
\end{equation}

\begin{equation}
\small
\label{maxscore}
\begin{aligned}
MaxScore = Max(sim(mbp_1, mbp_2 \in MBP_2)
\end{aligned}
\end{equation}

For each MBP in $MBP_1$, the highest similarity score, denoted as $MaxScore$, is first obtained using Equation (\ref{maxscore}) in $MBP_2$. Then, the similarity score of two MBP sets $sim(MBP_1, MBP_2)$ is the sum of the highest similarity score of each MBP in $MBP_1$ timed by the length of $mbp_1$ and divided by a base, which is the total length of all MBPs in $MBP_1$. This allows the similarity score to vary between zero and one, inclusively.

The similarities between functions are not sufficient to describe the similarity relationship between the target program and the candidate program. We combine the matched function pairs and the function innovation relationship into a graph that is the similar part of the function level birthmarks, which can reconstruct the similarity scene. Consequently, to prove the similarity between the target program and the candidate program, the similar parts between their three-level birthmarks are distilled to serve as the interpretable detection results. The similarity between the target program and candidate program can be measured according to the following Equation ~(\ref{simPro})
\begin{equation}
\small
\label{simPro}
\begin{aligned}
& sim(T, C) =  \frac{|FF|}{|C|}&
\end{aligned}
\end{equation}

The similarity score of the target program and candidate program $sim(T, C)$  is calculated by using the number of matched function pairs $FF$ to divide a base, which is the function number of candidate program. The score denotes the percent of reused functions in the candidate program with regard to the target program.

\section{Evaluation}
	\label{Sec_Evaluation}
In this section, we first introduce the setup of our experiments. Then, we answer the following six research questions to validate the performance of our approach.  

\newcounter{myrq}
\setcounter{myrq}{1}
\renewcommand\themyrq{\arabic{myrq}}
\renewcommand\thequestion{\themyrq}

\begin{question}
	\label{RQ-1}
	\textit{Can \textit{ISRD} effectively and efficiently detect partial reuse?}
\end{question}
\addtocounter{myrq}{1}
\begin{question}
	\label{RQ-2}
	\textit{Are the detection results of \textit{ISRD} interpretable?}
\end{question}
\addtocounter{myrq}{1}
\begin{question}
	\label{RQ-3}
	\textit{Is \textit{ISRD} resilient to cross-compiler-version?}
\end{question}
\addtocounter{myrq}{1}
\begin{question} 
	\label{RQ-4}
	\textit{Is \textit{ISRD} resilient to  cross-optimization-level ?}
\end{question}
\addtocounter{myrq}{1}
\begin{question}
	\label{RQ-5}
	\textit{Is \textit{ISRD} resilient to cross-compiler-vendor?}
\end{question}
\addtocounter{myrq}{1}
\begin{question}
	\label{RQ-6}
	\textit{How good is the result of \textit{ISRD}, compared to other related works?}
\end{question}
\addtocounter{myrq}{1}

%\vspace{-15pt}

\subsection{Study Setup}
\subsubsection{Evaluation Datasets}	
\textit{ISRD} takes the binary code of program pairs as input. To perform a thorough evaluation, we needed ground-truth datasets of which the binary code really share the same code. To this end, we used two datasets, including a dataset that is constructed by ourselves~(\textbf{Dataset-}\uppercase\expandafter{\romannumeral1}) and a widely used benchmark dataset \textbf{Dataset-}\uppercase\expandafter{\romannumeral2} that is provided from Bingo~\cite{chandramohan2016bingo} and $\alpha$diff~\cite{liu2018alphadiff}. 

To test whether \textit{ISRD} can successfully detect partial reuses, we collected real-world data from open source platforms to construct \textit{Dataset-}\uppercase\expandafter{\romannumeral1}. We first downloaded 24 open-source projects from open-source platforms~(e.g., Github and SourceForge), which fall into different application domains. Then, we selected and labelled a total of 74 real partial reuses as ground truth by using both a manual method and an automatic method. Finally, a dataset containing 24 programs with 74 partial reuses was constructed. We compiled these 24 programs into binaries using  \textit{gcc} with default optimization (O2). The static information of \textit{Dataset} \uppercase\expandafter{\romannumeral1} is listed in Table \ref{Tab-Sec4-Dataset1}, where the numbers in the third and the fourth columns are the lines of program source code and the number of functions in the corresponding compiled binaries, respectively. After compiling, some programs would generate more than one binary. Herein, we only present the information of the binaries that are used in our evaluation. We have published \textit{Dataset} \uppercase\expandafter{\romannumeral1} at~\cite{Dataset}.

\textbf{Dataset-}\uppercase\expandafter{\romannumeral2} is a widely-used benchmark to evaluate cross-compilation robustness. It was commonly used in the literatures~\cite{chandramohan2016bingo,liu2018alphadiff,egele2014blanket,wang2017memory}. Thus, it was adopted to compare the performance of \textit{ISRD} with existing approaches. 
The binaries in \textbf{Dataset-}\uppercase\expandafter{\romannumeral2} were compiled from the experimental object \textit{Coreutils}~\cite{Coreutils},  which are the basic file, shell and text manipulation utilities of the GNU operating system written in C. There are 107 components in \textit{Coreutils} and as a result, each compilation produces 107 binaries. 
Following \cite{chandramohan2016bingo,liu2018alphadiff,david2017similarity,david2016statistical}, we used three compilers in our experiment (\textit{gcc v4.6}, \textit{gcc v4.8}, and \textit{clang v3.0} ) and four optimization levels (O\{$0$, $1$, $2$, and $3$\}), resulting in 12 different variants for each of the 107 binaries.  Table \ref{Tab-Sec4-Dataset1} lists the information of \textbf{Dataset-}\uppercase\expandafter{\romannumeral2}, where the numbers in the 4th-7th columns denote the maximum, minimum, average, and total numbers of functions of the 107 binaries in \textit{Coreutils}, respectively.

\begin{table}[t]
	\centering
	\vspace{-5pt}
	\scriptsize
	\caption{Descriptions of \textit{Dataset} \uppercase\expandafter{\romannumeral1}.}
	\scalebox{1}
	{
		%\vspace{-3pt}
		\begin{tabular}{p{2cm}<{\centering}p{1cm}<{\centering}p{1cm}<{\centering}p{1.1 cm}<{\centering}}
			\toprule[1.5pt]
			Program  & Version  &   \# LOC   &   \# Function \\ \midrule
			\multirow{2}{*} {bzip2~\cite{bzip2}}   & 1.0.6   & 6019 & 84\\
			& 1.0.8 & 6026 & 84\\
			\multirow{2}{*} {zstd~\cite{zstd}}   & 1.4.3   &  76465 &848 \\
		       & 1.4.5 &  78667  &895   \\
			\multirow{2}{*} {lzo~\cite{lzo}}   & 2.09   &  23736 & 206\\
			                       & 2.10 &  24002  & 208  \\        
			minizip~\cite{minizip} & 2.8.7 & 28478 &  577 \\
		    precomp~\cite{precomp} & 0.4.7 & 96840 & 1739 \\
			TurboBench~\cite{TurboBench}& -& 509182  & 2486 \\
			lzbench~\cite{lzbench}& 1.8 &207315 & 2664 \\
			brotli~\cite{brotli} &1.0.7&  30072 & 233\\
			libbsc~\cite{libbsc}& 3.1.0 &  9192 & 70\\
			libdeflate~\cite{libdeflate} &1.6& 79071 & 70\\
			lzfse~\cite{lzfse} &1.0& 3382& 35\\
			lzlib~\cite{lzlib} &1.11& 4709 & 98\\
			zlib~\cite{zlib} &1.2.11&25504 & 133 \\
			zlib-ng~\cite{zlib-ng} &-& 14288 &187\\
			csc~\cite{CSC}&-& 6373 & 119\\
			gipfeli~\cite{gipfeli}&-&1112 & 84\\
			blosc ~\cite{blosc}& 1.18.1&58867 & 742\\
			liblzg~\cite{liblzg}&1.0.10&1648 & 30\\
			xz~\cite{xz}&5.2.4& 24175 & 401\\
			Exserver~\cite{Exserver}&-& 5625 & 133\\
			cknit~\cite{cknit}&-& 6054 & 139\\
			exjson~\cite{exjson}&-& 3384 & 83\\
			libsndfile~\cite{libsndfile}&1.0.28& 50764 &44438\\
			sndfile2k~\cite{sndfile2k}&-&59532 &39299\\
			\toprule[1.5pt]
		\end{tabular}	
	}

	\label{Tab-Sec4-Dataset2}
% 	\vspace{-15pt}
\end{table}

\begin{table}[t]
	\centering
% 	\vspace{-5pt}
	\scriptsize
	\caption{Descriptions of \textit{Dataset} \uppercase\expandafter{\romannumeral2}.}
	\scalebox{1}
	{
		%\vspace{-3pt}
		\begin{tabular}{p{0.8cm}<{\centering}p{0.6cm}<{\centering}p{1.3cm}<{\centering}p{0.8cm}<{\centering}p{0.8cm}<{\centering}p{0.8cm}<{\centering}p{0.8cm}<{\centering}}
			\toprule[1.5pt]
		 	Compiler & Version & \tabincell{c}{Optimization \\ Level}  &
	 Max   &   Min  &  Average   &  Total  \\ \midrule
			\multirow{8}{*} {\textit{gcc}} &  \multirow{4}{*} {v4.6} 
		      & O0 & 379 & 17 & 128 & 13711 \\  
			& & O1 & 284 & 14 & 106 & 11382 \\  
			& & O2 & 275 &  14 & 110 & 11809 \\  
			& & O3 & 258 & 14 & 107 & 11510  
\\ \cmidrule{2-7}
			&  \multirow{4}{*} {v4.8} 
			  & O0 & 380 & 18 & 129 & 13825 \\  
			& & O1 & 266 & 15 & 104 & 11186 \\
			& & O2 & 277 & 15 & 112 & 12047 \\  
			& & O3 & 253 & 15 & 106 & 11424 \\   \midrule
			\multirow{4}{*} {\textit{clang}} &  \multirow{4}{*} {v3.0} 
			  & O0 & 484 & 17 & 168 & 17978 \\  
			& & O1 & 486 & 17 & 169 & 18110 \\  
			& & O2 & 278 & 13 & 116 & 12504 \\  
			& & O3 & 267 & 13 & 115 & 12308 \\  
			
			\toprule[1.5pt]
		\end{tabular}	
	}
	\label{Tab-Sec4-Dataset1}
	%\vspace{-15pt}
\end{table}

\subsubsection{Evaluation Metrics}
The metrics used to measure the performance of \textit{ISRD} are shown in TABLE \ref{Tab-Sec4-Evaluation Metrics}.
\textit{False Positive Rate} (FPR) stands for the ratio of non-reusable functions being falsely detected as reusable functions. \textit{False Negative Rate} (FNR) quantifies the ratio of the reusable functions that are not detected as reusable functions. The values of \textit{precision}, \textit{Recall} and \textit{F-Measure} are calculated as described in TABLE \ref{Tab-Sec4-Evaluation Metrics}.  %In reuse detection, \textit{precision} is calculated by dividing the number of correctly predicted samples by the total number of positively predicted samples, which refers to the percentage of accurately detected reuses in a dataset. \textit{Recall} is measured by taking the number of correctly predicted samples divided by the total number of positive samples. \textit{F-Measure} is the harmonic average of \textit{precision} and \textit{recall} .

\begin{table}[t]
	\centering
	\vspace{-5pt}
	\scriptsize
	\caption{Evaluation Metrics.}
	\scalebox{1}
	{
		%\vspace{-3pt}
		\begin{tabular}{p{2.5cm}<{\centering}p{0.6cm}<{\centering}p{4.5 cm}<{\centering}}
			\toprule[1.5pt]
			Term            & Abbr  &  Definition \\ \midrule
			\multirow{2}{*} {True Positive}   & \multirow{2}{*} {TP }    &  the number of reusable functions that are \\
			                      & &        correctly detected as reusable. \\
			\multirow{2}{*} {True Negative}   & \multirow{2}{*} {TN}    &  the number of unreusable functions that are \\
			& & correctly detected as unreusable. \\
			\multirow{2}{*} {False Negative}  & \multirow{2}{*} {FN}    &  the number of reusable functions that are \\
			& &incorrectly detected as unreusable. \\ 
			\multirow{2}{*} {False Positive}  & \multirow{2}{*} {FP}    &  the  number of unreusable functions that are\\
			& & incorrectly detected as reusable. \\
			False Positive Rate & FPR & $ FP/(FP+TN) $\\
			False Negative Rate & FNR & $FN/(TP+FN)$\\
		    Precision  & P   &  $ TP/(TP+FP) $  \\
			Recall & R    & $ TP/(TP+FN) $ \\
			F-measure  & F$_1$ &  $2PR/(P + R) $ \\
			\toprule[1.5pt]
		\end{tabular}	
	}
	\label{Tab-Sec4-Evaluation Metrics}
	\vspace{-15pt}
\end{table}	

\subsubsection{Parameter Setting and Experimental Environment}
Similarity threshold plays an important role in \textit{ISRD}. A function pair is determined as matched if its similarity score is greater than the similarity threshold. To determine a proper threshold, we varied its values to test over both datasets. According to our testing results, setting the similarity threshold as 0.5 made \textit{ISRD} achieve the best performance.

We implemented \textit{ISRD} in \textit{python 3.6} on \textit{Ubuntu 18.04}. All programs ran at DELL desktop P2417H with CPU i7-7700 \& 3.60GHz and 16GB memory. In the experiments, \textit{ISRD} utilized \textit{angr}~\cite{angr} to disassemble binaries. 

\subsection{Answer to RQ~\ref{RQ-1}: Effectiveness and Efficiency of \textit{ISRD}}
\begin{figure}[t]
	\centering	
	\subfloat[CDF of Precision]{\includegraphics[width=0.23\textwidth]{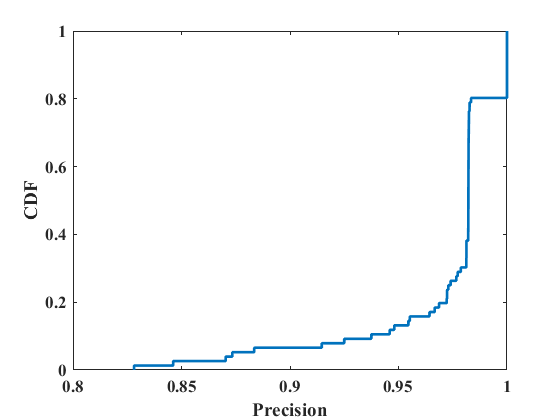}}
	\hspace{3mm}
	\subfloat[CDF of Recall]{\includegraphics[width=0.23\textwidth]{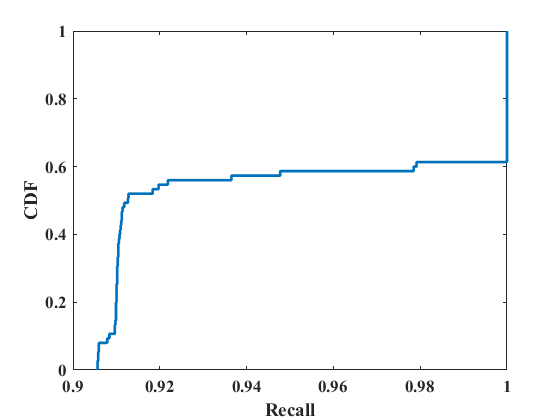}}
	
	\caption{CDFs of \textit{ISRD} Precision and Recall on Dataset~\uppercase\expandafter{\romannumeral1} }
	\label{Fig-Sec3-RQ1-1}
	\vspace{-8pt}
\end{figure}

\begin{figure}[t]
	\centering
	\includegraphics[width=0.23\textwidth]{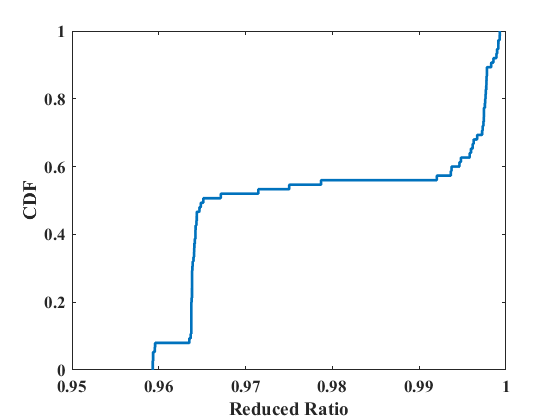}
	\centering
	\caption{CDF for Reduced Ratio of Partial Reuse Detection on Dataset~\uppercase\expandafter{\romannumeral1}}
	\label{Fig-Sec3-RQ1-2}
	\vspace{-15pt}
\end{figure}

In this evaluation, we used \textit{Dataset} \uppercase\expandafter{\romannumeral1} to test whether \textit{ISRD} can effectively and efficiently detect partial reuse. We ran \textit{ISRD} with \textit{Dataset} \uppercase\expandafter{\romannumeral1} as input and compared its results with the ground truth. Fig.~\ref{Fig-Sec3-RQ1-1} and Fig.~\ref{Fig-Sec3-RQ1-2} report the performance of \textit{ISRD} in terms of effectiveness and efficiency to detect partial reuse. 

Fig.~\ref{Fig-Sec3-RQ1-1} presents the Cumulative Distribution Function~(CDF) for the precision and recall of the evaluation results. Almost all the precision values are higher than $82\%$ and its average value can reach $97.2\%$. Moreover,  15 binary pairs achieve precision equal to 1, indicating that all reused functions can be accurately detected as reusable. For the recall metric, its average value reaches $94.8\%$, indicating that our approach can effectively find reused function pairs. 

%Precision is achieved $97.8\%$ on average and the average recall reaches $88.3\%$. Moreover, 

Fig.~\ref{Fig-Sec3-RQ1-2} reports the CDF with regard to the ratio of the reduced number of function pairs that were calculated by \textit{ISRD} to the original number of function pairs. We can observe that \textit{ISRD} effectively reduces the size of the function pairs by $97.9\%$ on average during reused function detection. Specifically, when the reused code is only a small fraction of the target program and the candidate program, the reduced ratio is up to $99.9 \%$. For example, the reused code of \textit{bzip2} only account for about $2\%$ of \textit{precomp}, 208824 functions pairs are needed to be calculated if applying some existing approaches (i.e., BinDiff~\cite{bindiff}, TRACY~\cite{david2014tracelet}), whereas 128 function pairs are calculated by \textit{ISRD}. Thus, we can conclude that the number of function pairs used in \textit{ISRD} is much smaller than the total number of all function pairs between the target program and the candidate program. This huge reduction in reuse detection makes \textit{ISRD} practical for large real-world programs.

%\vspace{2pt}
\begin{tcolorbox}
\textbf{Answering RQ~\ref{RQ-1}:}  
\textit{ISRD} effectively detect partial reuse and its average precision achieves $97.2\%$. \textit{ISRD} can significantly reduce the number of function pairs required for reuse detection up to $97.9\%$ on average. Thus, it is efficient to handle a large scale of programs.
\end{tcolorbox}

\subsection{Answer to RQ~\ref{RQ-2}: Interpretability of \textit{ISRD}}

To evaluate whether the detection results of \textit{ISRD} are interpretable, we leveraged two programs  \textit{precomp}~\cite{precomp} and \textit{minizip}~\cite{minizip} in \textit{Dataset} \uppercase\expandafter{\romannumeral1} to perform a case study. The case study illustrates the interpretability of ISRD by providing a graphic demonstration of reuse. Herein, \textit{precomp} is a command line precompressor and \textit{minizip} is a zip manipulation library in C. Note that both \textit{precomp}  and \textit{minizip} reuse two compression libraries \textit{lzma}~\cite{lzma} and \textit{bzip2}~\cite{bzip2}. 

\begin{figure*}[t]
	\centering
	\includegraphics[width=0.85\textwidth]{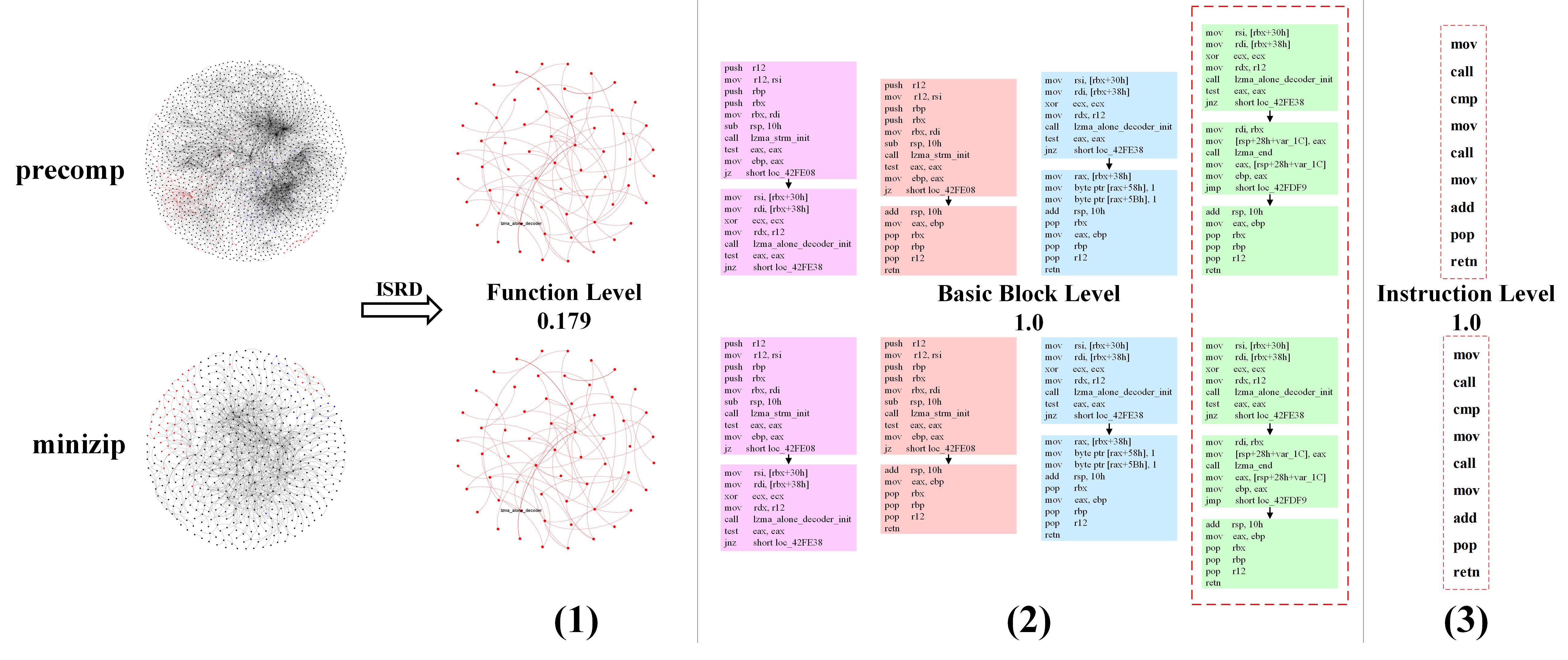}
	\centering
	\caption{The Detection Results of \textit{minizip} and \textit{precomp}}
	\label{Fig-Sec4-CaseStudy}
	\vspace{-15pt}
\end{figure*}

Fig.~\ref{Fig-Sec4-CaseStudy} illustrates the detection results of \textit{ISRD}. The FCGs of \textit{precomp}  and  \textit{minizip} are presented on the left side of the figure,
where the nodes in red and in blue represent the functions in the library \textit{lzma} and \textit{bzip2}, respectively. Running \textit{ISRD} on  \textit{precomp} and  \textit{minizip} , we got program similarity as $17.9 \%$, which implied that $17.9\%$ functions of \textit{precomp} have been already used in \textit{minizip}.

The detailed detection results are illustrated and interpreted on the right side of the figure. For presentation purpose, only the \textit{lzma} is given. After identifying all matched function pairs between \textit{precomp} and  \textit{minizip}, the matched subgraph were constructed by combining the matched function pairs with their function call relationship in order to reconstruct the reuse scene in the function level of the two programs. The matched subgraph between the two programs is presented in Fig.~\ref{Fig-Sec4-CaseStudy}~(1). In the matched subgraph, the node pairs are the matched function pairs between the two programs that have the same function call relationship.

Furthermore, we singled out function \textit{lzma\_alone\_decoder} with its matched function, whose similarity score is 1.0,  as an example by including the matched part of this function at the basic block level and the instruction level. In the basic block level, as showed in Fig.~\ref{Fig-Sec4-CaseStudy}~(2), 4 matched MBP pairs were recognized to demonstrate the similarity between the function pair. The matched MBPs are displayed with the same colors.

Finally, the similarity relationship of the last matched MBP pair is described in detail at the instruction level. 9 high-level matched operation pairs explain the semantic equivalence at the finest granularity, as displayed in Fig.~\ref{Fig-Sec4-CaseStudy}~(3) with a similarity score as 1.0.

%\vspace{2pt}
\begin{tcolorbox}
\textbf{Answering RQ~\ref{RQ-2}:}  
The case study shows that the detection results of \textit{ISRD} is interpretable by describing the matched part between the target program and the candidate program in detail at the function level, the basic block level and the instruction level.
\end{tcolorbox}
%\vspace{-12}

\subsection{Answer to RQ~\ref{RQ-3}: Resilience to Cross-Compiler-Version}
We compiled \textit{Coreutils} using \textit{gcc v4.6} and \textit{gcc v4.8} with various optimization levels (O0 to O3). Such a setup led to 8 different versions for each binary in \textit{Coreutils}. Subsequently, we evaluated \textit{ISRD} by comparing the binaries compiled using different compiler versions  with the same optimization levels.

Table \ref{Tab-Sec4-CrossCompilerVersion} summarizes this experiment's results, where each row heading represents the optimization level used for compilation, and the last row reports the average value of each evaluation metric. For example, the second row represents the detection results when the target programs and the candidate programs were compiled using \textit{gcc v4.6} and \textit{gcc v4.8} with the same optimization level, O0.

\begin{table}[t]
	\centering
	\scriptsize
	\caption{Detection Results on Resilience to Cross-Compiler-Version}
	\scalebox{0.95}{
		%\vspace{-3pt}
		\begin{tabular}{p{1.6cm}<{\centering}p{0.8cm}<{\centering}p{0.8cm}<{\centering}p{0.8cm}<{\centering}p{0.8cm}<{\centering}p{0.8cm}<{\centering}}
			
	\toprule[1.5pt]
    gcc 4.6-gcc 4.8  &  P  &  R  & F$_1$  &  FPR  &  FNR  \\ \midrule
    O0-O0 & 1.000 & 0.996 & 0.998 & 0.000 & 0.004  \\ 
    O1-O1 & 0.997 & 0.990 & 0.994 & 0.000 & 0.010  \\ 
    O2-O2 & 0.911 & 0.997 & 0.952 & 0.003 & 0.003  \\ 
    O3-O3 & 0.932 & 0.987 & 0.958 & 0.001 & 0.013  \\ \midrule
    Average & 0.960 & 0.993 & 0.975 & 0.001 & 0.007  \\ 

			\bottomrule[1.5pt]
		\end{tabular}	
	}
	%\vspace{-10pt}
	\label{Tab-Sec4-CrossCompilerVersion}
	\vspace{-15pt}
\end{table}	

The results demonstrate that the average precision can achieve $96.0\%$ and the recall is $99.3\%$, indicating that \textit{ISRD} is resilient to the obfuscation caused by different compiler versions. Moreover, for no code optimization (i.e., O0), the precision is $100\%$ while the FPR is $0\%$. That is, even with different version compilers, the compilations without code optimization levels lead to highly similar binary codes,  whereas compiling with high optimization levels yields more differences between binaries.

%\vspace{2pt}
\begin{tcolorbox}
\textbf{Answering RQ~\ref{RQ-3}:}  
\textit{ISRD} is resilient to cross-compiler-version with $96.0\%$  precision and $99.3\%$ recall on average.
\end{tcolorbox}
\vspace{-3pt}

\subsection{Answer to RQ~\ref{RQ-4}: Resilience to Cross-Optimization-Level}
We compiled \textit{Coreutil} for x86-32bit architecture using \textit{clang v3.0} and \textit{gcc v4.8} with various optimization levels (O0 to O3). With this setup, each binary in \textit{Coreutils}  had 8 variants. We compared the binaries compiled using the same compiler with different optimization levels.

\begin{table}[t]
	\centering
	\scriptsize
	\caption{Detection Results on Resilience to Cross-Optimization-Level}
	\scalebox{0.95}{
		%\vspace{-3pt}
		\begin{tabular}{p{0.6cm}<{\centering}p{1cm}<{\centering}p{0.8cm}<{\centering}p{0.8cm}<{\centering}p{0.8cm}<{\centering}p{0.8cm}<{\centering}p{0.8cm}<{\centering}}
			
			\toprule[1.5pt]
  &  &  P  &  R  & F$_1$  &  FPR  &  FNR  \\  \midrule
  \multirow{7}{*} {gcc 4.8}  
  & O0-O1 & 0.977 & 0.753 & 0.847 & 0.000 & 0.247  \\ 
& O0-O2 & 0.859 & 0.764 & 0.806 & 0.003 & 0.236  \\ 
& O0-O3 & 0.863 & 0.624 & 0.720 & 0.001 & 0.376  \\ 
& O1-O2 & 0.873 & 0.982 & 0.924 & 0.004 & 0.018  \\ 
& O1-O3 & 0.913 & 0.923 & 0.916 & 0.001 & 0.077  \\ 
& O2-O3 & 0.980 & 0.973 & 0.976 & 0.000 & 0.027  \\  \cmidrule{2-7}
   &Average & 0.911 & 0.836 & 0.865 & 0.002 & 0.164  \\    \midrule

 \multirow{7}{*} {clang 3.0 }   
 & O0-O1 & 0.968 & 0.975 & 0.971 & 0.001 & 0.025  \\ 
&O0-O2 & 0.962 & 0.792 & 0.864 & 0.002 & 0.208  \\  
&O0-O3 & 0.960 & 0.784 & 0.858 & 0.002 & 0.216  \\ 
&O1-O2 & 0.929 & 0.844 & 0.883 & 0.003 & 0.156  \\ 
&O1-O3 & 0.925 & 0.835 & 0.876 & 0.003 & 0.165  \\ 
&O2-O3 & 1.000 & 0.999 & 0.999 & 0.000 & 0.001  \\  \cmidrule{2-7}
&Average & 0.957 & 0.871 & 0.909 & 0.002 & 0.129  \\  
			\bottomrule[1.5pt]
		\end{tabular}	

	}
	%\vspace{-10pt}
	\label{Tab-Sec4-CrossOptimizationLevel}
	\vspace{-15pt}
\end{table}	

Table \ref{Tab-Sec4-CrossOptimizationLevel} summarizes the results, where each row heading represents the optimization level used to compile the detection objects. For example, the second row represents the detection results of the target programs that were compiled by \textit{gcc v4.8} with O0 while the candidate programs were compiled with O1 using the same compiler. 

We observe that \textit{ISRD} performs much better regarding precision in \textit{clang v3.0} than in \textit{gcc v4.8}. Specifically, the average precision is $95.7\%$ for \textit{clang v3.0} yet only $91.1\%$ is obtained for \textit{gcc v4.8}.

Another interesting observation is that we get similar patterns for both compilers. Within one compiler type (e.g., \textit{gcc v4.8}.), the F-measure achieved by matching between the binaries that were all compiled with high code optimization levels (i.e., O2 and O3) is always better than that obtained by matching between the binaries where one is compiled with a high code optimization level (i.e., O2 and O3) and the other is compiled with no code optimization (i.e., O0). Specifically, the highest F-measure is achieved when the target program and the candidate program were both compiled with high optimization, O2 and O3, respectively. However, F-measure drops to $72.0\%$ if the target program and the candidate program were respectively compiled with no code optimization (i.e., O0) and high code optimization level (i.e., O3). This suggests that regardless of the compiler vendor, the binaries compiled with high optimization levels are similar. The influence of whether the optimization kicks into the binaries is very evident. 

%\vspace{2pt}
\begin{tcolorbox}
\textbf{Answering RQ~\ref{RQ-4}:}  
\textit{ISRD} is resilient to cross-optimization-level to some extent. Evaluating the binaries compiled by \textit{gcc v4.8} and \textit{clang v3.0}, \textit{ISRD} achieves on average  $91.1\%$  precision and $95.7\%$ precision, respectively.
\end{tcolorbox}
\vspace{-3pt}

\subsection{Answer to RQ~\ref{RQ-5}: Resilience to Cross-Compiler-Vendor}
We compiled \textit{Coreutil} for x86-32bit architecture using \textit{clang v3.0} and \textit{gcc v4.8} with various optimization levels (O0 to O3). 8 different variants were generated for each binary in \textit{Coreutils} with this setup. Subsequently, we evaluated \textit{ISRD} on binaries compiled by different vendors' compilers with same optimization levels.

We list the results in Table \ref{Tab-Sec4-CrossCompilerVendor}, where each row heading represents the optimization level used to compile the compared objects. %For example, the third row represents the detection result of the target programs and the candidate programs that were compiled using \textit{clang v3.0} and \textit{gcc v4.8} with the same optimization level, O1. 
Experimental results show that the average precision reaches a high degree, $94.2\%$ with FPR as $0.1\%$. From the table, we find that the best results are obtained when the candidate programs were compiled with no optimization level as the target one. Besides, the precision hits the lowest point of $91.2\%$ when the detection binaries  were compiled with optimization level O3. Further, across compiler vendors, the detection precision drops with the rise of the optimization level, except when the detection programs were compiled with optimization level O1, where the detection programs compiled with optimization level O2 yield better accuracy.  .

\begin{table}[t]
	\centering
	\scriptsize
	\caption{ Detection Results on Resilience to Cross-Compiler-Vendor }
	\scalebox{0.95}{
		%\vspace{-3pt}
		\begin{tabular}{p{1.6cm}<{\centering}p{0.8cm}<{\centering}p{0.8cm}<{\centering}p{0.8cm}<{\centering}p{0.8cm}<{\centering}p{0.8cm}<{\centering}}
			
			\toprule[1.5pt]
			
   clang3.0-gcc4.8  & P  &  R  & F$_1$  &  FPR  &  FNR \\  \midrule
   O0-O0 & 0.999 & 0.919 & 0.956 & 0.000 & 0.081  \\ 
O1-O1 & 0.929 & 0.765 & 0.835 & 0.001 & 0.235  \\ 
O2-O2 & 0.930 & 0.903 & 0.913 & 0.002 & 0.097  \\ 
O3-O3 & 0.912 & 0.867 & 0.886 & 0.001 & 0.133  \\ \midrule
Average & 0.942 & 0.864 & 0.898 & 0.001 & 0.136  \\

			\bottomrule[1.5pt]
		\end{tabular}	
	}
	%\vspace{-10pt}
	\label{Tab-Sec4-CrossCompilerVendor}
	\vspace{-5pt}
\end{table}

% \vspace{2pt}
\begin{tcolorbox}
\textbf{Answering RQ~\ref{RQ-5}:}  
Experimental results demonstrate the resilience of \textit{ISRD} across compiler vendors. Moreover, it achieves on average  $94.2\%$ precision.
\end{tcolorbox}
\vspace{-3pt}

\subsection{Answer to RQ~\ref{RQ-6}: Comparison with Related Works}
In this experiment, we compared \textit{ISRD} with three baseline approaches, including, BinDiff~\cite{bindiff}, BinGo~\cite{chandramohan2016bingo}, and $\alpha$Diff~\cite{liu2018alphadiff}. %These approaches are briefly described below:

\begin{itemize}
\item[1)] Bindiff~\cite{bindiff} is a binary code similarity detection tool, which matches a pair of binaries using a variant of graph-isomorphism algorithm.
\item[2)] Bingo~\cite{chandramohan2016bingo} is a scalable and robust binary search engine that supports cross-compilation by applying a selective inlining technique to recover complete function semantics. 
\item[3)] $\alpha$diff~\cite{liu2018alphadiff} is a method to detect binary code similarity with a neural network solution to extract intra-function semantic features from raw bytes of binary functions.
% \item[4)] BinGo-E~\cite{xue2019accurate} is a binary code search approach, which leverages on emulation to extract low-level semantic features. 
\end{itemize}

To make a head-to-head comparison with these approaches, we used the same experimental configurations and the same metric as theirs. More specifically, we compiled \textit{Coreutils} using \textit{gcc v4.8} and \textit{clang v3.0} with various optimization levels (from O0 to O3). We evaluated six experimental settings. The recall values of reuse detection are reported in Table~\ref{Tab-Sec4-Comparison}, where each row heading represents the optimization level used to compile the target programs and the candidate programs. %For example, the second row represents that the target programs were compiled by \textit{clang v3.0} with the optimization level O0 whereas the candidate programs were compiled by \textit{gcc v4.8} with the optimization level O3.

\begin{table}[t]
	\centering
	\scriptsize
	\caption{Comparison with Three Baseline Approaches Indicated by Recall}
	\scalebox{0.95}{
		%\vspace{-3pt}
		\begin{tabular}{p{2.4cm}<{\centering}p{0.6cm}<{\centering}p{0.6cm}<{\centering}p{0.6cm}<{\centering}p{0.6cm}<{\centering}}
			
			\toprule[1.5pt]
                &  Bindiff  &  BinGo  &  $\alpha$diff    &  ISRD  \\  \midrule
    clang-O0 vs. gcc-O3 & 0.271     & 0.332   & 0.462            &  0.644  \\
    clang-O0 vs. clang-O3 & 0.351     & 0.372   & 0.492            &  0.784 \\
    clang-O2 vs. clang-O3 & 0.994     & 0.576   & 0.969           &  0.999\\
    gcc-O0 vs. clang-O3 & 0.258     & 0.333   & 0.484            &  0.562 \\
    gcc-O0 vs. gcc-O3 & 0.255     & 0.302   & 0.441             &  0.624\\
    gcc-O2 vs. gcc-O3 & 0.757     & 0.480   & 0.765             &  0.973\\ \midrule
    Average & 0.481 & 0.399 & 0.602 & 0.764 \\

			\bottomrule[1.5pt]
		\end{tabular}	
	}
	%\vspace{-10pt}
	\label{Tab-Sec4-Comparison}
	\vspace{-15pt}
\end{table}	

Through comparison on recall, we can see that \textit{ISRD} outperforms the three baseline approaches by $27.0\%$ on average, especially outperforms BinGo by $36.5\%$ on average. We note that the average recall of \textit{ISRD} is $76.4\%$, which is lower than the upper experiment results. This is because the obfuscations of both cross-compiler-vendor and cross-optimization-level were involved, making the detection much harder than the situation where only one type of obfuscation occurs. Moreover, the worst result was obtained by each approach when the target programs were compiled by \textit{gcc v4.8} with no code optimization and the candidate programs were compiled by \textit{gcc v4.8} with the highest optimization level O3. On the other hand, the best results of all approaches were achieved when the target programs and the candidate programs were compiled by the same compiler with the high optimization level, O2 and O3, respectively. 

%\vspace{2pt}
\begin{tcolorbox}
\textbf{Answering RQ~\ref{RQ-6}:}  
Compared with the baseline approaches, \textit{ISRD} achieves better performance regarding a widely used benchmark dataset.
\end{tcolorbox}
\vspace{-3pt}

\section{Limitations of ISRD}
\label{Sec_Limitation}
\subsection{Impacts Caused by Internal Reasons}
\textbf{Function Inlining.} Function inlining is a major internal limitation to \textit{ISRD}. In \textit{ISRD}, the function semantics inside a program is distilled independently. To be specific, a callee function's semantics is not integrated into a caller function's semantics. This leads to a partial semantics problem, because the strategy of inlining is selectively applied according to a configured optimization level during compilation. Inlining is utilized in compilers to optimize the binaries for achieving maximum speed or minimum size ~\cite{chang1992profile-guided}. Other approaches (e.g., the selective inlining strategy of Bingo ~\cite{chandramohan2016bingo}) that inline relevant libraries and developer-defined function semantics can be incorporated into \textit{ISRD} to address the problem.

\textbf{Library Functions.} In \textit{ISRD}, two ways are leveraged to recognize the anchors. The second way is based on checking identical library call invocations. However, this way is limited for two reasons:~(1) the library functions are OS dependent;~(2) it fails to recognize the library calls that have different names yet with similar functionality (e.g.,\textit{memcpy} and \textit{memmove})~\cite{chandramohan2016bingo}. To address the above problems, inspired by CLCDSA~\cite{nafi2019clcdsa}, the similarity of cross-os library calls can be learned with the help of the documentation and Mikolov's Word2Vec~\cite{mikolov2013distributed} model. 

\subsection{Impacts Related to Datasets}
Due to the difficulty of collecting partial reuse samples with accurate labels,  only 24 programs were selected to construct \textit{Dataset} \uppercase\expandafter{\romannumeral1} and 74 real partial reuses were manually labeled. In future work, we plan to collect additional real-world partial reuse samples and further evaluate and optimize the performance of \textit{ISRD}.

\section{Conclusion}
	\label{Sec_Conclusion}
In this paper, we proposed \textit{ISRD}, an interpretation-enabled software reuse detection approach based on a multi-level birthmark model. We used the multi-level birthmark model to distill the program semantics from coarse granularity to fine granularity. Through normalization, the function semantics and the core semantics of instructions can be effectively represented. In reuse detection, intent search originated from recognized anchors was employed to speed up function pair matching. Extensive experiments reveal that \textit{ISRD} is effective and efficient in detecting partial reuse with interpretation. It also outperforms state-of-the-art approaches in terms of resilience to cross-compilation.

\section*{Acknowledgment}

This work was supported by National Key R\&D Program of China (2018YFB1004500), National Natural Science Foundation of China (61632015, 61772408, U1766215, 61721002, 61532015, 61833015, 61902306, 62072351), China Postdoctoral Science Foundation (2019TQ0251, 2020M673439), Youth Talent Support Plan of Xi'an Association for Science and Technology (095920201303), Ministry of Education Innovation Research Team (IRT\_17R86), and Project of China Knowledge Centre for Engineering Science and Technology, as well as Academy of Finland (Grants 308087 and 335262).

\bibliographystyle{IEEEtran}
\bibliography{mybibfile}

\end{document}